\documentclass{article}
\usepackage{dcolumn}
\usepackage{bm}
\usepackage{latexsym}
\usepackage{amsfonts}
\usepackage{amssymb}
\usepackage{amsmath}
\def\abstract#1{\vskip 7mm 
        \begin{center}{\large Abstract}\par \smallskip
                \begin{minipage}[c]{12cm}
                        \small #1
                \end{minipage}
        \end{center}
}
\def\title#1{\begin{center}{\Large\bf #1}\end{center}}
\def\author#1{\vskip 5mm \begin{center}{#1}\end{center}}
\def\address#1{\begin{center}{\it #1}\end{center}}
\makeatletter
\@ifundefined{lesssim}{}{}
\@ifundefined{gtrsim}{}{}
\def\vereq#1#2{\lower3pt\vbox{\baselineskip1.5pt \lineskip1.5pt
\ialign{$\m@th#1\hfill##\hfil$\crcr#2\crcr\sim\crcr}}}
\makeatother

\begin{document}

\title{%
Impact of Lorentz Violation on Cosmology
}
\author{%
  Jiro Soda$^1$
  and
  Sugumi Kanno$^2$
}
\address{%
  $^1$Department of Physics,  Kyoto University, Kyoto 606-8501, Japan\\
  $^2$Department of Physics,  McGill University, Montr${\acute e}$al, 
 QC H3A 2T8, Canada
}

\abstract{
 We discuss the impact of Lorentz violation on the cosmology.
 Firstly, we show that the Lorentz violation  affects the dynamics of the
 chaotic inflationary model and gives rise to an interesting feature. 
 Secondly, we propose the Lorentz violating DGP brane models 
 where the Lorentz violating terms on the brane accelerate the current universe. 
 We conjecture that the ghost disappears in the Lorentz violating DGP models.
}

\section{Introduction}

Various observations suggest the existence of two accelerating stages in the universe,
namely, the past inflationary universe and the current acceleration of the universe.
In cosmology, therefore, how to accelerate the universe in the past and present
is a big issue. Many attempts to resolve this issue have been performed and failed.
Given the difficulty of the problem, it would be useful to go back to
the basic point and reexamine it. Here, we consider the possibility
to break the Lorentz symmetry.

Typically, the Lorentz violation yields the preferred frame.
In the case of the standard model of particles, 
there are strong constraints on the existence of the preferred frame. 
In contrast, there is no reason to refuse the preferred frame in cosmology.
Rather, there is a natural preferred frame
defined by the cosmic microwave background radiation (CMB).
Therefore, there is room to consider the gravitational theory
which allows the preferred frame. 

Now, we present our model with which we discuss
the cosmological acceleration problems. 
Suppose that the Lorentz symmetry is spontaneously
broken by getting the expectation values of a vector field $u^\mu$ as
$
     <0| u^\mu u_\mu |0> = -1 \ .
$
We do not notice the existence of this field because the frame
determined by this field coincides with the CMB frame.
However, in the inhomogeneous universe, both frames can fluctuate independently.
Hence, we can regard the spatial hypersurface of our universe as a kind of
membrane characterized by the extrinsic curvature $K_{ij}$ 
with a time like vector field $u^\mu$. Based on this observation, we propose the model
\begin{eqnarray}
  S &=& \int d^4 x \sqrt{-g} \left[ \frac{1}{16\pi G} R
  +\beta_1 (\phi) K^{ij} K_{ij} - \beta_2 (\phi) K^2
  - \gamma_1 (\phi) \nabla^\mu u^\nu \nabla_\mu u_\nu \right. \nonumber\\
&& \left.  \qquad\qquad\qquad
   -\gamma_2 (\phi )\nabla^\mu u^\nu \nabla_\nu u_\mu    
   -\gamma_3 (\phi) \left( \nabla_\mu u^\mu \right)^2 
  -\gamma_4 (\phi) u^\mu u^\nu \nabla_\mu u^\alpha \nabla_\nu u_\alpha 
                                   \right. \nonumber \\
&& \left.  \qquad\quad\quad\qquad\qquad
     + \lambda \left( u^\mu u_\mu +1 \right)
  - \frac{1}{2} \left(\nabla \phi \right)^2 - V(\phi )
                             \right] \ ,
\end{eqnarray}
where $g_{\mu\nu}$, $R$, and $\lambda$ are the 4-dimensional metric,
 the scalar curvature, and a Lagrange multiplier, respectively.
Here, we also considered the scalar field $\phi$ with the potential $V$
and it couples to other terms with the coupling function $\beta_i (\phi)$ and $\gamma_i (\phi)$. 
This is a generalization of the Einstein-Ether gravity~\cite{Jacobson}.
 Since $\beta_i , \gamma_i$ at present can be different from
those in the very early universe, we do not have any constraint on 
these parameters in the inflationary stage. Of course, ultimately, they
have to approach the observationally allowed values at present.
 Here, the Lorentz symmetry is violated both spontaneously and 
 explicitly. 
 
 The purpose of this paper is to discuss the impact of the Lorentz
 violation both on the inflationary scenario and the current
 acceleration. In the former case, the Lorentz violation merely modifies 
 the scenario.
 In the latter case, however, the impact of Lorentz violation could be appreciable.
 The inclusion of Lorentz violation may give rise to a resolution of the ghost problem
 in the DGP brane model~\cite{Dvali:2000hr}.

\section{Impact of Lorentz Violation on Inflationary Scenario}

Now, let us consider the chaotic inflationary scenario and clarify 
to what extent the Lorentz violation affects the inflationary
scenario~\cite{Kanno:2006ty}. We take the model
\begin{eqnarray}
  S = \int d^4 x \sqrt{-g} \left[\frac{1}{16\pi G} R -  \beta (\phi) K^2
   - \frac{1}{2} \left(\nabla \phi \right)^2 - V(\phi ) \right] \ .
\end{eqnarray}
Let us consider the homogeneous and isotropic spacetime
\begin{eqnarray}
ds^2 = -  dt^2 + e^{2\alpha(t)} \delta_{ij} dx^i dx^j \ . 
\end{eqnarray}
 The scale of the universe is determined by $\alpha$. 
 Now, let us deduce the equations of motion.
First, we define the dimensionless derivative $Q'$ by
$
  \dot{Q}  =  \frac{dQ}{d\alpha} \frac{d\alpha }{dt} 
            \equiv Q' \frac{d\alpha }{dt}\ .
$ 
Then, the equations of motion are 
\begin{eqnarray}
&&  \left( 1 + \frac{1}{8\pi G \beta} \right) H^2 
  = \frac{1}{3} \left[
  \frac{1}{2} \frac{H^2 \phi^{\prime 2}}{\beta} + \frac{V}{\beta} 
                      \right] \\
&&   \left( 1 + \frac{1}{8\pi G \beta} \right)\frac{H'}{H} 
     + \frac{1}{2} \frac{\phi^{\prime 2}}{\beta} + \frac{\beta'}{\beta} =0 \\
&&  \phi'' + \frac{H'}{H} \phi' + 3\phi' + \frac{V_{,\phi}}{H^2}
             + 3 \beta_{,\phi} =0             \ ,        
\end{eqnarray}
where $\beta_{,\phi}$ denotes the derivative with respect to $\phi$.
We have taken  $H=\dot{\alpha}$ as an independent variable.
As is usual with gravity, these three equations are not independent.
Usually, the second one is regarded as a redundant equation.

The above equation changes its property at the critical value $\phi_c$
defined by
$
8\pi G \beta (\phi_c ) =1  \ .
$
When we consider the inflationary scenario, 
 we usually require the enough e-folding number, say $N=70$. Let $\phi_i$
be the corresponding initial value of the scalar field.
If $\phi_c > \phi_i$, the effect of Lorentz violation on the inflationary
scenario would be negligible. 
However, if $\phi_c < \phi_i$, the standard scenario
should be modified. It depends on the models. 
To make the discussion more specific, we choose the model
$
  \beta = \xi \phi^2 \ , \quad V = \frac{1}{2} m^2 \phi^2 \ ,
$
where $\xi$ and $m$ are parameters. 
For this model, we have 
$
   \phi_c = \frac{M_{pl}}{\sqrt{8\pi \xi}}  \ .
$
As $\phi_i \sim 3 M_{pl}$ approximately in the standard case,
 the condition $\phi_i > \phi_c$ implies the criterion
$\xi > 1/(72\pi) \sim 1/226$ for the Lorentz violation to be relevant 
to the inflation. For other models, the similar criterion can
be easily obtained. 

Now, we suppose the Lorentz violation is relevant and
analyze the two regimes separately.

For a sufficiently larger value of $\phi$, both the coupling function
$\beta$ and the potential function $V$ are important in the model (2). 
During this period, the effect of Lorentz violation on the
inflaton dynamics must be large. 
In the Lorentz violating regime, $8\pi G \beta \gg 1$, we have
\begin{eqnarray}
&&   H^2 
  = \frac{1}{3\beta} \left[
  \frac{1}{2} H^2 \phi^{\prime 2} + V 
                      \right] \\
 &&   \frac{H'}{H} 
     + \frac{1}{2\beta} \phi^{\prime 2} + \frac{\beta'}{\beta} =0 \\         
&&  \phi'' + \frac{H'}{H} \phi' + 3\phi' + \frac{V_{,\phi}}{H^2}
             + 3 \beta_{,\phi} =0     \ .                
\end{eqnarray}
To have the inflation, we impose the condition 
$
H^2 \phi^{\prime 2} \ll V
$
as the slow roll condition. Consequently, Eq.(7) is reduced to
\begin{eqnarray}
   H^2   = \frac{1}{3\beta} V   \ .                
\end{eqnarray}
Using Eq.(10), the slow roll condition  can be written as
$
\phi^{\prime 2} \ll \beta  \ .
$
Now, we also impose the condition $H'/H \ll 1$
 as the quasi-de Sitter condition.
Then, Eq.(8) gives us the condition
$
  \beta' \ll \beta \ .
$
We also require the standard condition
$
   \phi'' \ll \phi' \ . 
$
Thus, we have the slow roll equations (10) and 
\begin{eqnarray}
  \phi' + \frac{V_{,\phi}}{3 H^2} +  \beta_{,\phi} =0     \ .                
\end{eqnarray}

For our example, we can easily solve Eqs.(10) and (11) as
$
   \phi (\alpha) = \phi_i e^{- 4\xi \alpha} \ .
$
For this solution to satisfy slow roll conditions, 
we need $ \xi <1/16$. Thus, we have the range $1/226 <\xi <1/16$
 of the parameter for which the Lorentz violating inflation is relevant.
 Note that, in our model, the Hubble parameter (10) becomes constant
\begin{eqnarray}
H^2 = \frac{m^2}{6\xi} \ ,
\end{eqnarray}
even though the inflaton is rolling down the potential.
This is a consequence of Lorentz violation. 

After the inflaton crosses the critical value $\phi_c$, the dynamics
is governed entirely by the potential $V$.  
In the standard slow roll regime $8\pi G \beta \ll 1$, 
the evolution of the inflaton can be solved as
\begin{eqnarray}
    \phi^2 (\alpha) = \phi_c^2 - \frac{\alpha}{2\pi G } \ .
\end{eqnarray}
The scale factor can be also obtained as
$
  a (t)  = \exp \left[ 2\pi G ( \phi_c^2 -\phi^2 (t) ) \right] \ .
$
The standard inflation stage ends and the reheating commences
when the slow roll conditions violate.

Now it is easy to calculate e-folding number.
Let $\phi_i$ be the value of the scalar field corresponding to
the e-folding number $N=70$. 
The total e-folding number reads
\begin{eqnarray}
  N = \frac{1}{4\xi} \log \frac{\phi_i}{\phi_c}
        + 2\pi G \left( \phi_c^2 -\phi_e^2 \right) \ ,
\end{eqnarray}
where $\phi_e \sim 0.3 M_{pl}$ is the value of  
scalar field at the end of inflation.
Note that the first term arises from the Lorentz violating stage.
As an example, let us take the value $\xi = 10^{-2}$. 
Then, $\phi_c \sim 2 M_{pl}$.
The contribution to the e-folding number from the inflation end is negligible.
Therefore, we get $\phi_i \sim 12 M_{pl}$. 

In this simple example, the coupling to 
the Lorentz violating sector disappears after 
the reheating. Hence, the subsequent homogeneous dynamics of the universe 
is the same as that of Lorentz invariant theory of gravity.
However, it is possible to add some 
 constants to $\beta$, which are consistent with the current
experiments. In that case, the effect of the Lorentz violation
is still relevant to the subsequent history.

The tensor part of perturbations can be described by
\begin{eqnarray}
ds^2 =  -dt^2 
    + a^2 (t)\left( \delta_{ij} + h_{ij} (t,x^i) \right) dx^i dx^j  \ ,
\end{eqnarray}
where the perturbation satisfy $h^i{}_i = h_{ij}{}^{,j}=0$.
The quadratic part of the action is given by
\begin{eqnarray}
  S = \int d^4 x \frac{a^3}{16\pi G} \left[
      \frac{1}{4} 
      \dot{h}_{ij} \dot{h}^{\prime ij} 
      -\frac{1}{4a^2} h_{ij,k} h^{ij,k} \right] \ .
\end{eqnarray}
In the case of chaotic inflation model, the Hubble parameter is constant
 (12) during Lorentz violating stage. The spectrum is completely
flat although the inflaton is rolling down the potential. 
This is a clear prediction of the Lorentz violating chaotic inflation.

\section{Impact of Lorentz Violation on Current Acceleration}

To solve the current acceleration problem is much more difficult
than the past one. The most interesting proposal is the DGP model~\cite{Dvali:2000hr}.
However, it suffers from the ghost. Hence, it is not a stable model.
Here, we would like to argue the Lorentz violation may resolve 
the instability problem.

Our basic observation is that 
the Lorentz violating term itself can accelerate the universe.
Let us consider the simplest braneworld model:
\begin{eqnarray}
  S = \frac{1}{2\kappa_5^2}\int d^5 x \sqrt{-G} R - \int d^4 x \sqrt{-g} \beta K^2 , 
\end{eqnarray}
where $G$ is the determinant of the 5-dimensional metric and $\kappa_5$
is the 5-dimensional gravitational coupling constant.
Here, we have assumed the scalar field is stabilized at present.
Let us assume the $Z_2$ symmetry. Then the junction condition
$
K_{\mu\nu}-g_{\mu\nu} K = \kappa_5^2 T_{\mu\nu}
$
gives the effective Friedman equation:
\begin{eqnarray}
  \pm H = \frac{\kappa_5^2}{6}\left[ 3 \beta H^2 -\rho \right] \ .
\end{eqnarray}
If we take the positive sign, in the late time, we have the de Sitter spacetime with 
\begin{eqnarray}
   H = \frac{2}{\kappa_5^2 \beta}  \ .
\end{eqnarray}
Thus, the late time accelerating universe can be realized 
due to the Lorentz violation. 

More generally, we propose the following model
\begin{eqnarray}
  S &=& \frac{1}{2\kappa_5^2}
           \int d^5 x \sqrt{-G} R \nonumber\\
&& \quad
  + \int d^4 x \sqrt{-g} \left[ \frac{1}{16\pi G} R
  +\beta_1  K^{ij} K_{ij} - \beta_2  K^2
  - \gamma_1  \nabla^\mu u^\nu \nabla_\mu u_\nu \right. \nonumber\\
&& \left.  \qquad\qquad\qquad\qquad
   -\gamma_2 \nabla^\mu u^\nu \nabla_\nu u_\mu    
   -\gamma_3  \left( \nabla_\mu u^\mu \right)^2 
  -\gamma_4  u^\mu u^\nu \nabla_\mu u^\alpha \nabla_\nu u_\alpha 
     + \lambda \left( u^\mu u_\mu +1 \right)
                             \right] \ .
\end{eqnarray}
It should be stressed that any term on the brane can generate the current acceleration.
As the tensor structure in the action is very different from the original DGP model,
 we can expect the above action contains no ghost. 
 In fact, we have many parameters here, hence there is a chance
to remove the ghost from our models. 
 Thus, we expect that some of the above Lorentz violating brane models
does not contain  any ghost.  If this is so,
one can say the Lorentz violation explains the current stage acceleration of 
the universe.

\section{Conclusion}

We have discussed the impact of Lorents violation on cosmology.
In the first place,
we found that the Lorentz violating inflation shows an interesting feature.
In the second place, we proposed a Lorentz violating DGP model
which have a possibility to avoid the ghost problem.

It would be interesting to study the evolution of fluctuations completely.
If the vector modes of perturbations
can survive till the last scattering surface, they leave the remnant
of the Lorentz violation on the CMB polarization spectrum. 
It is also intriguing to seek for a relation to the large scale
anomaly discovered in CMB 
by WMAP.
The calculation of the curvature perturbation is much more complicated.
However, it must reveal more interesting phenomena due to
Lorentz violating inflation. The tensor-scalar
ratio of the power spectrum would be also interesting. 
These are now under investigation.  

More importantly, we need to show the stability of Lorentz violating DGP model. 
If the Lorentz violation kills the ghost, this is a great progress in the cosmology.
Even in case that it turns out that all of the models
are unstable, it makes our understanding of the ghost issue profound.

\vskip 0.5cm
S.K. is supported by JSPS Postdoctoral Fellowships
for Research Abroad. 
J.S. is supported by the Grant-in-Aid for the 21st Century COE "Center for Diversity and Universality in Physics" from the Ministry of Education, Culture, Sports, Science and Technology (MEXT) of Japan, 
the Japan-U.K. Research Cooperative Program, the Japan-France Research
Cooperative Program,  Grant-in-Aid for  Scientific
Research Fund of the Ministry of Education, Science and Culture of Japan 
 No.18540262 and No.17340075.

\end{document}